# Polarized lines of the second solar spectrum (SrI, SrII, CaI, BaII) observed at the Pic du Midi Turret Dome spectropolarimeter with the slit orthogonal to the limb

*Jean-Marie Malherbe (emeritus astronomer)*

Observatoire de Paris, PSL Research University, LIRA, France

Email: Jean-Marie.Malherbe@obspm.fr; ORCID: https://orcid.org/0000-0002-4180-3729

30 May 2026

**ABSTRACT**

The second solar spectrum is the spectrum of the Stokes parameter Q (linear polarization) close to the solar limb. It differs significantly of the usual intensity spectrum (Stokes parameter I). The second solar spectrum contains in the blue just a few polarized lines with Q/I of about 1% (such as CaI, SrI, SrII, BaII, most lines exhibit much weaker polarization. This paper presents new processing of observations made in 2004-2006 with the Pic du Midi Turret Dome spectropolarimeter, which are of strong interest for weak and turbulent unresolved magnetic field measurements in the quiet Sun, through the interpretation of the Hanle effect. As the slit was orthogonal to the limb, the polarization rate Q/I is determined precisely and continuously up to 80" distance of the limb. All data shown in this paper are now available in the on-line FITS dataset for further investigations at: https://entrepot.recherche.data.gouv.fr/dataset.xhtml?persistentId=doi:10.57745/DF54CS.

## I - INTRODUCTION

Scattering processes on the Sun are sources of polarization in the solar spectrum (Stenflo & Keller, 1997; Stenflo, 2004). The polarization is zero at disk centre (symmetry of the radiation field around the line of sight) and increases towards the solar limb (where the line of sight is orthogonal to the incident radiation field of the photosphere). The Stokes parameter Q represents linear polarization parallel to the limb, and the ratio Q/I is the polarization rate. To determine this polarization, the slit of our solar spectrograph was orthogonal to the limb, in the vicinity of the solar equator, and the camera integrated for a long time to reach sufficient signal to noise ratio (SNR). Just a few lines exhibit about 1% polarization, such as SrI 4607 A, SrII 4078 A, BaII 4554 A or CaI 4227 A lines in the blue part of the spectrum. Instrumental polarization may be an obstacle for such observations, but in our case, we used the 50 cm refractor of the Pic du Midi turret dome (equatorial mount described by Roudier *et al*, 2021) with the polarimeter (Malherbe *et al*, 2007a, 2007b) at the primary focus, without any reflection, which guarantees almost no parasitic polarization. After passing through the polarimeter, the beam was injected into the 8 m spectrograph (Mouradian, 1980), a Littrow design of 8 m focal length. The dispersive element was an echelle grating (300 grooves/mm, 63°26' blaze angle) which offers about 10 mA spectral resolution in the blue part of the spectrum.

SrI, SrII, BaII or CaI lines are of great interest for the measurements of weak (tens of Gauss) and turbulent (unresolved) magnetic fields of the quiet Sun owing to the Hanle effect (Stenflo, 1982).

In the presence of magnetic fields, one observes a depolarization and a rotation of the polarization plane; however our observations do not include Stokes U, so that they are useful only for the depolarization, but not for the rotation.

## II - OBSERVATIONS WITH THE PIC DU MIDI SPECTROPOLARIMETER

The polarimeter is located at the F1 focus of the refractor at F/13 and is described by Malherbe *et al* (2007a, 2007b). It is made of two Liquid Crystal Variable Retarders (LCVR) and a precision dichroic linear polarizer from Meadowlark company (single beam polarimetry). Then the beam is injected into the spectrograph by a flat mirror at 45°, before magnification by a lens (equivalent focal length from 30 m to 60 m according to the magnification). The polarimeter can run at 50 Hz but is limited by the speed of the camera (5-10 Hz) and exposure time. It has no chromatism, because the voltage is adapted to the observed line to produce exactly zero, quarter or half wave retardance.

The polarimeter delivered the combinations I+Q and I-Q in sequence (only one LCVR was mounted). The driving voltages were chosen for each line so that the LCVR was strictly modulating between zero and half wave. The camera was a classical interline CCD camera with micro lenses (SONY sensor with 6.5 microns pixel) and electronic shutter. As the full well capacity is only 18000 electrons, corresponding to 12 bits (4.5 electrons/count), we alternated at 10 Hz cadence 16 exposures of I+Q and I-Q (16 couples) which were added in real time before FITS file writing. Hence we registered full 16 bit data corresponding to a full well capacity of 288 000 electrons, which corresponds to a SNR of about 500. Several tens or hundreds of observations obtained within a loop were then summed in order to reach a final SNR of about 10000 for typically 400 accumulations of 16 bit spectra. The spectral resolution was around 10 mA, depending on the line. The spectral pixel along the slit was 0.2 arcsec, and the length of the slit was 160”.

The refractor has no stabilizating image system, this has little importance at a few arcsec from the limb (statistical effect of summing a large amount of exposures) but it may affect measures at the limb itself where there is a steep intensity gradient. As the spectrograph is attached to the equatorial mount, its position varies with time, and this effect may generate line shifts; to avoid this, lines were tracked by the data processing software before curvature and inclination correction.

Observations were performed in the vicinity of the equator (slit orthogonal to the limb) and in a few cases near the north or south poles (slit parallel to the limb). Dark currents were subtracted and Flat Fields (FF) were measured at disk centre. FF are useful to correct the transmission between the two states of the LCVR delivering alternatively I+Q and I-Q. Flat Fields were observed systematically at disk centre in the quiet Sun (where the linear polarization should be zero) after each limb sequence. Unfortunately, in some cases, the FF was not usable because a parasitic wavelength shift occurred between the limb and disk centre; the absence of good FF inhibits the determination of the continuum polarization, but has a little impact on polarized lines such as those observed here. In such a case, we assumed the polarization to be null in the close continuum of lines and in average along the slit (which covers a field of view of 160” from the limb towards disk centre direction).

## III – POLARIZATION OF LINES AS A FUNCTION OF LIMB DISTANCE

We did several runs for BaII 4554 A, SrI 4607 A, SrII 4078 A and CaI 4227 A which are the most polarized lines of the second solar spectrum and the most useful for the Hanle effect (however, the derivation of weak turbulent magnetic fields through this effect remains outside the scope of this observational paper).

### III – 1 – BaII 4554 A (slit orthogonal to the limb near equator)

Main results are reported in Figures 1 to 7; all scientific data, and more than those of the figures, are included in the FITS dataset (three dates of observations, with slit orthogonal to the limb). BaII D2 is a good diagnostic tool for weak magnetic fields (Faurobert *et al*, 2009).

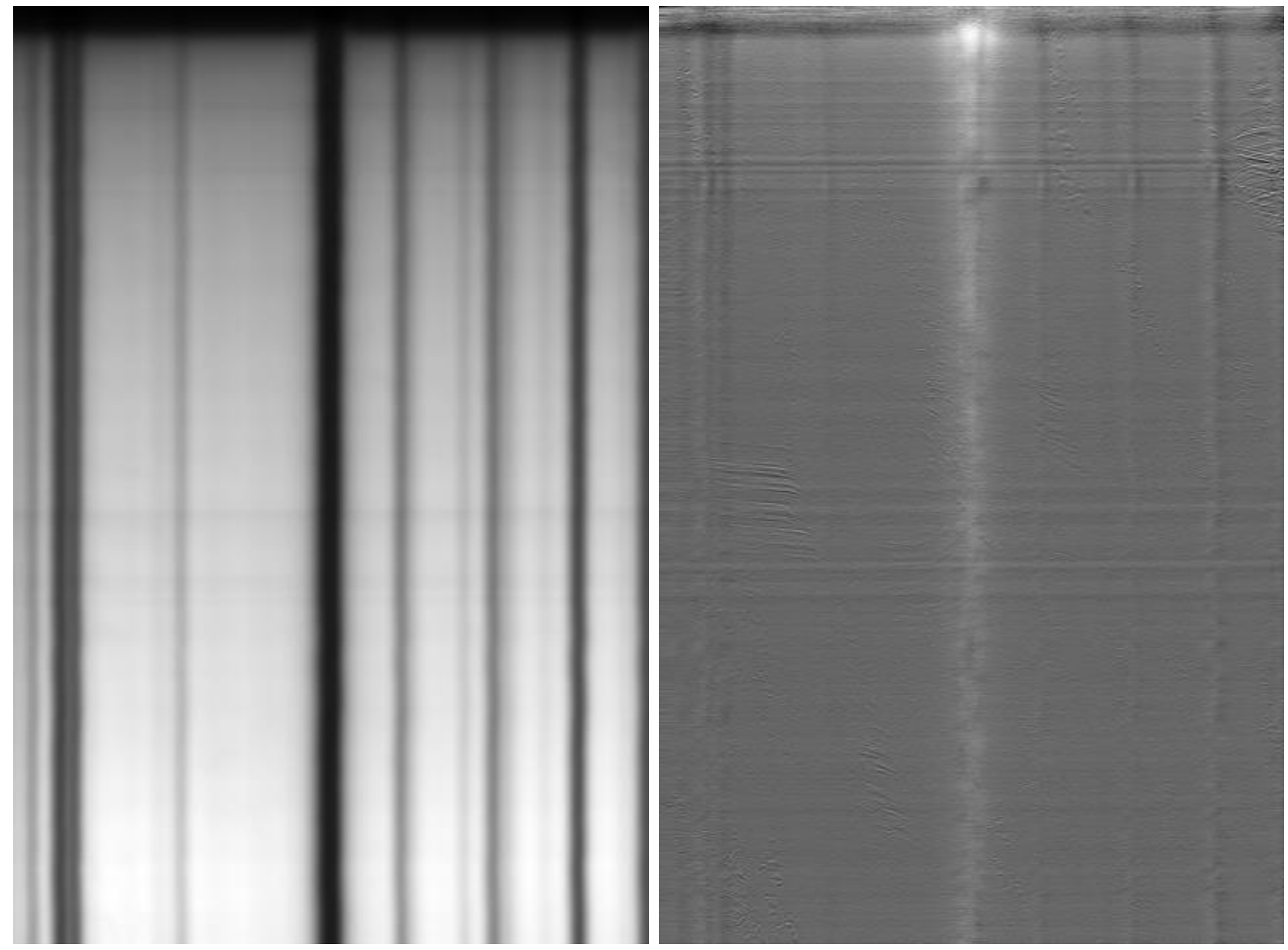

*Figure 1 : Intensity I(λ, x) and polarization rate Q/I(λ, x) where λ is the wavelength (in abscissa) and x (in ordinates) is the abscissa along the slit. 11 mA/pixel and 0.2''/pixel. 14 May 2004, 15:23 – 16:17 UT. The limb is at the top.*

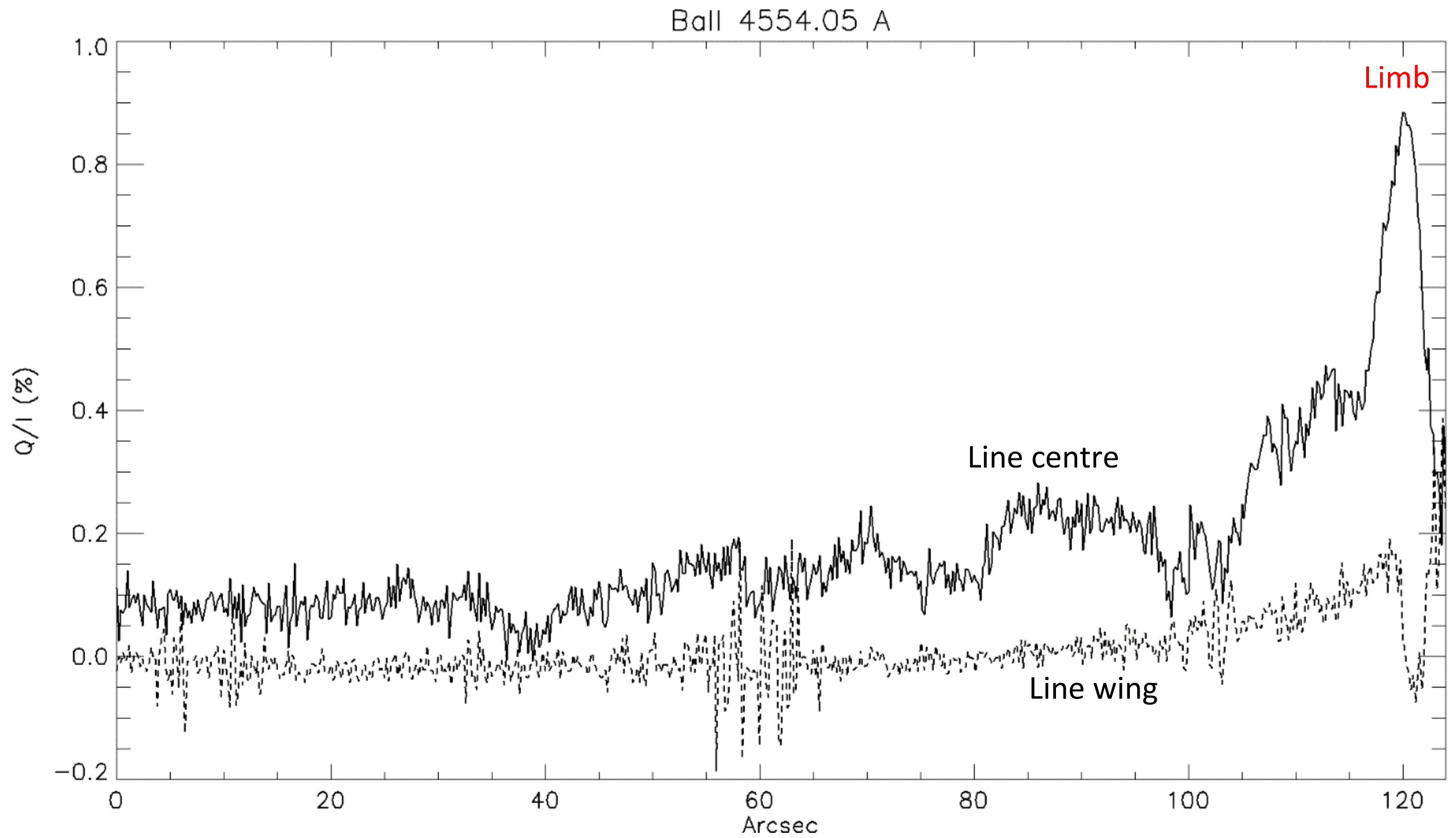

*Figure 2 : polarization rate Q/I(x) along the slit (the limb is at the abscissa x = 120") at line centre and in the blue continuum close to the line. 14 May 2004, 15:23 – 16:17 UT*

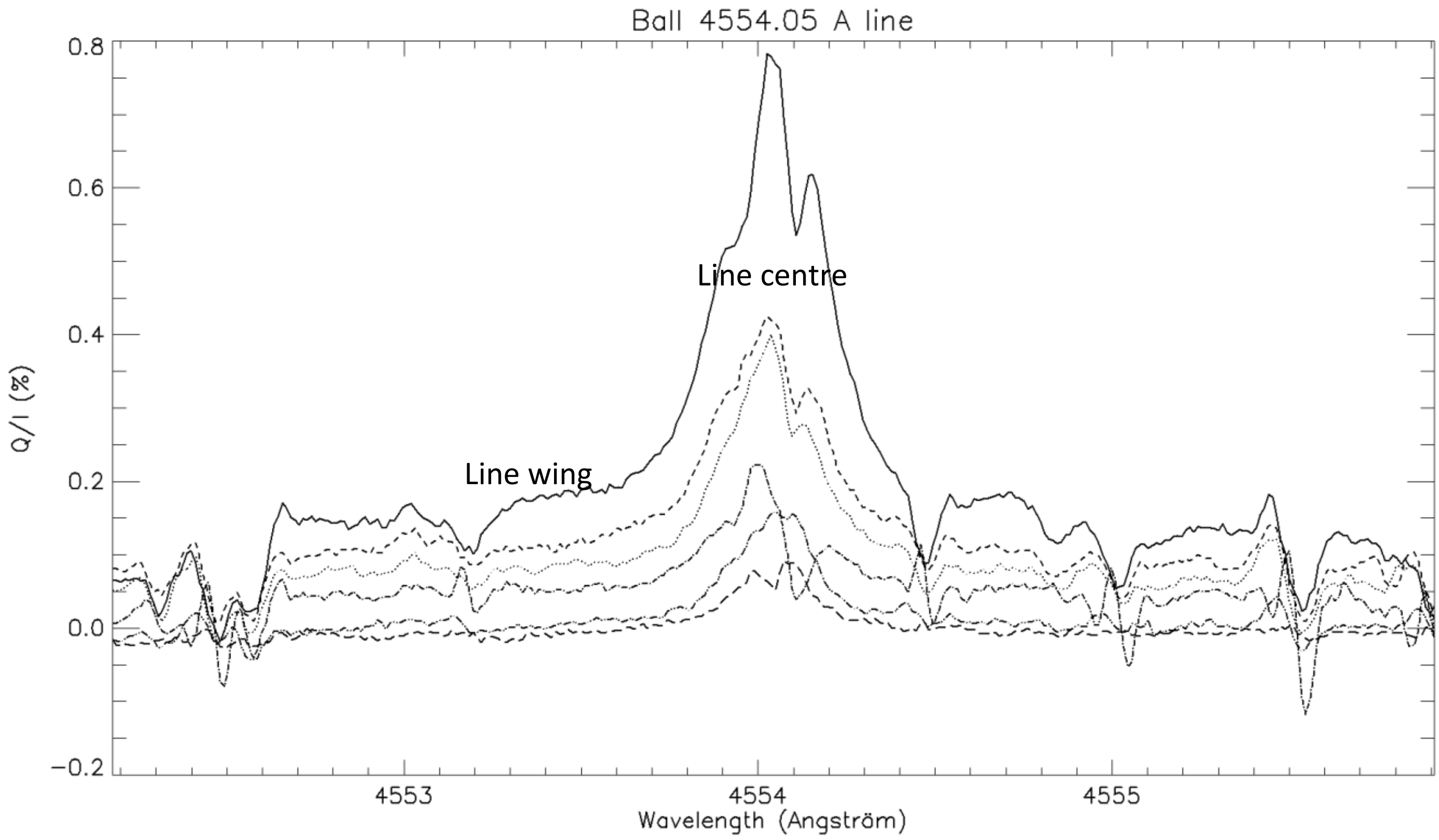

*Figure 3 : Q/I(λ) profiles (in %) at various limb distances of the limb (2", 5", 10", 20", 40", 80"). 14 May 2004, 15:23 – 16:17 UT.*

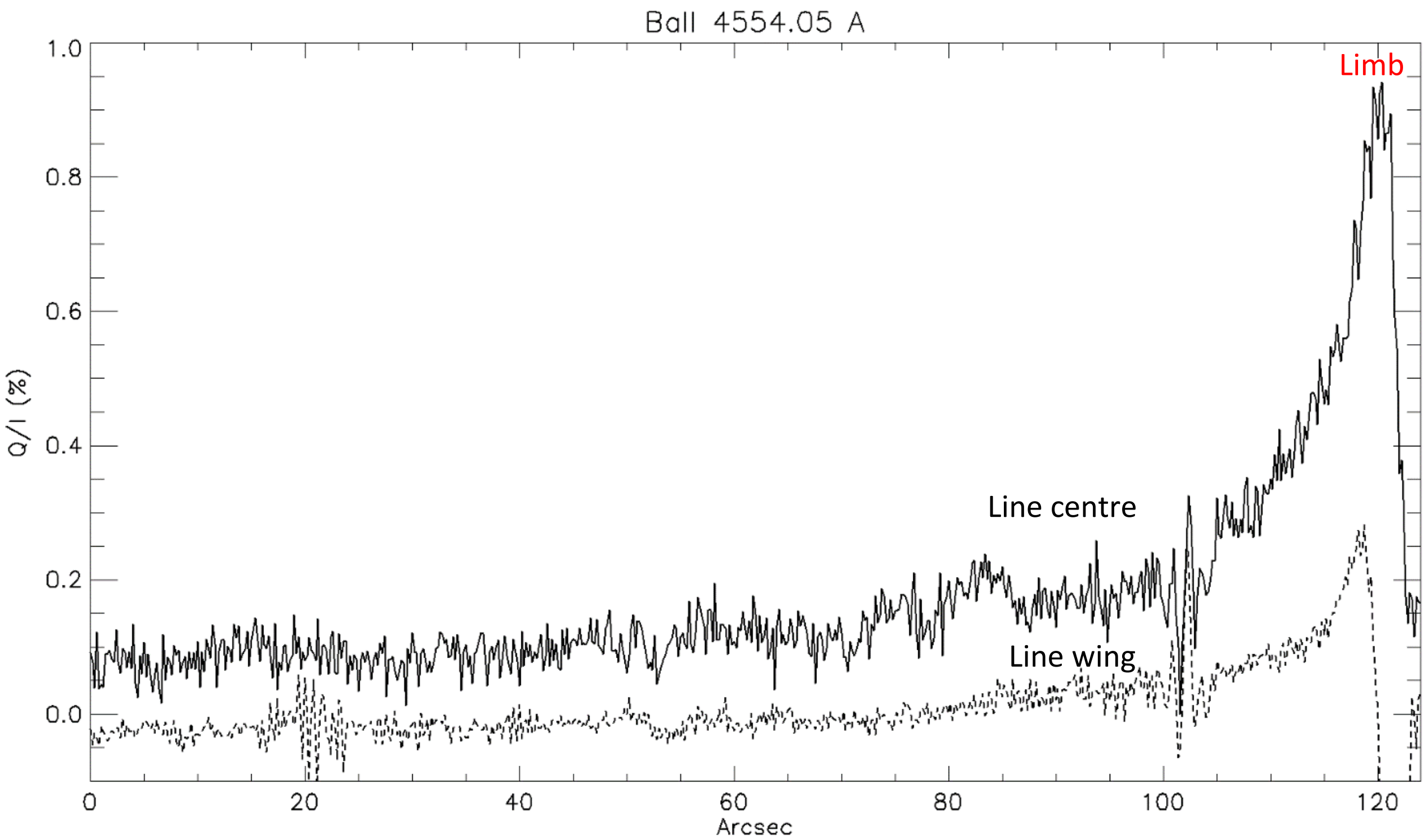

*Figure 4 : polarization rate Q/I(x) along the slit (the limb is at the abscissa x = 120") at line centre and in the blue continuum close to the line. 15 May 2004, 09:14 – 10:18 UT.*

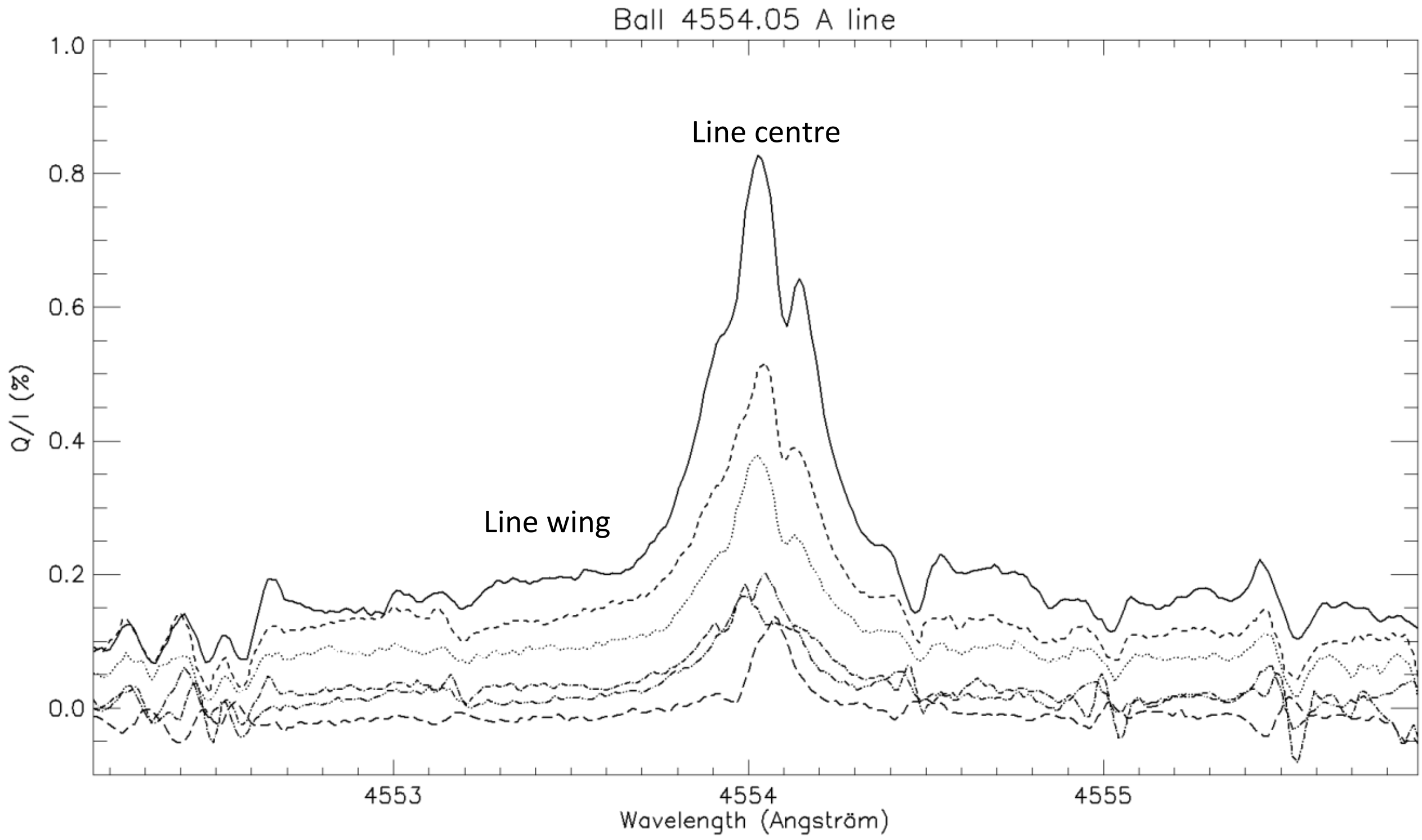

*Figure 5 : Q/I(λ) profiles (in %) at various limb distances of the limb (2", 5", 10", 20", 40", 80"). 15 May 2004, 09:14 – 10:18 UT.*

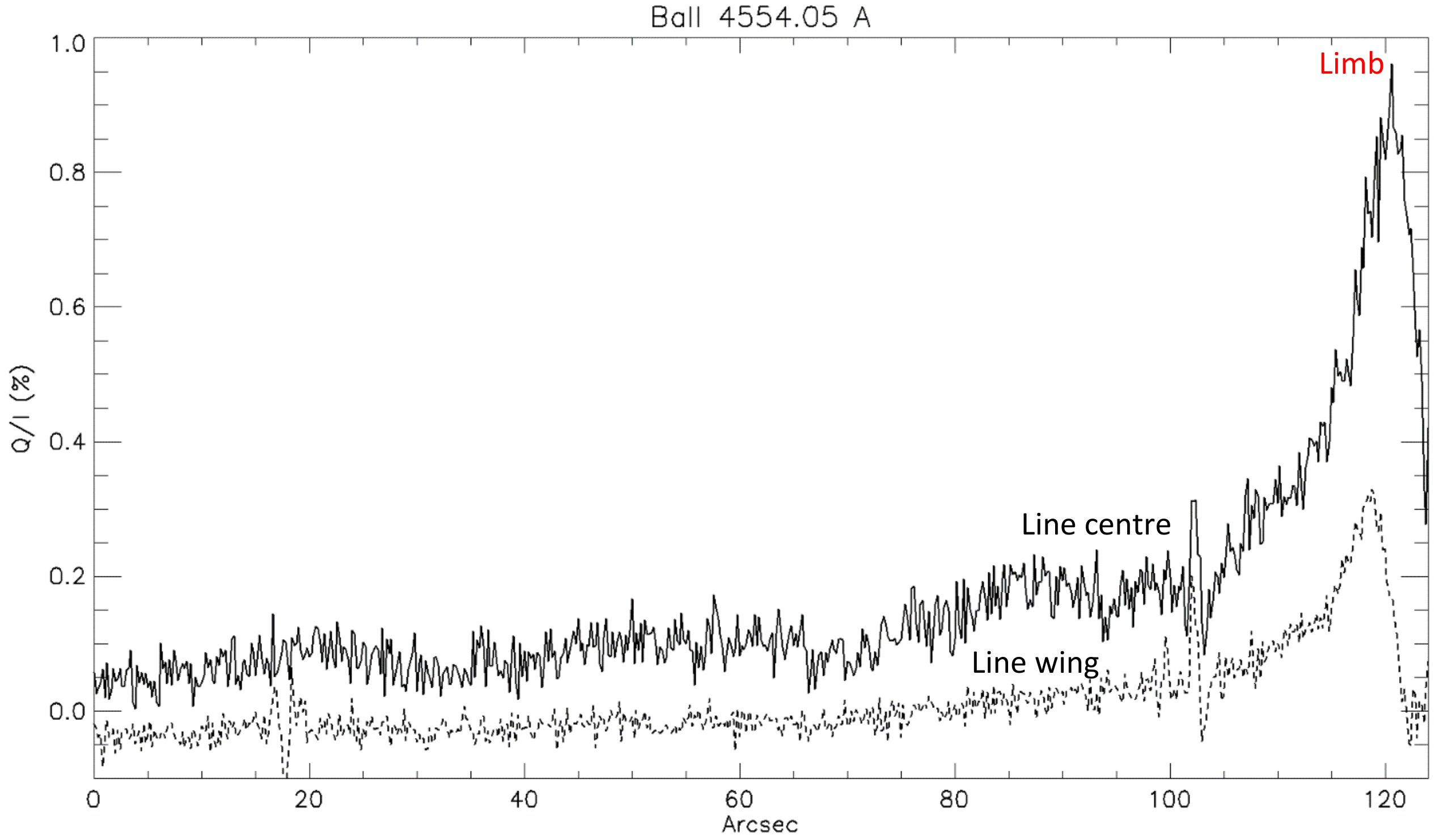

*Figure 6 : polarization rate Q/I(x) along the slit (the limb is at the abscissa x = 120") at line centre and in the blue continuum close to the line. 15 May 2004, 09:14 – 10:18 UT. 15 May 2004, 10:20 – 11:20 UT.*

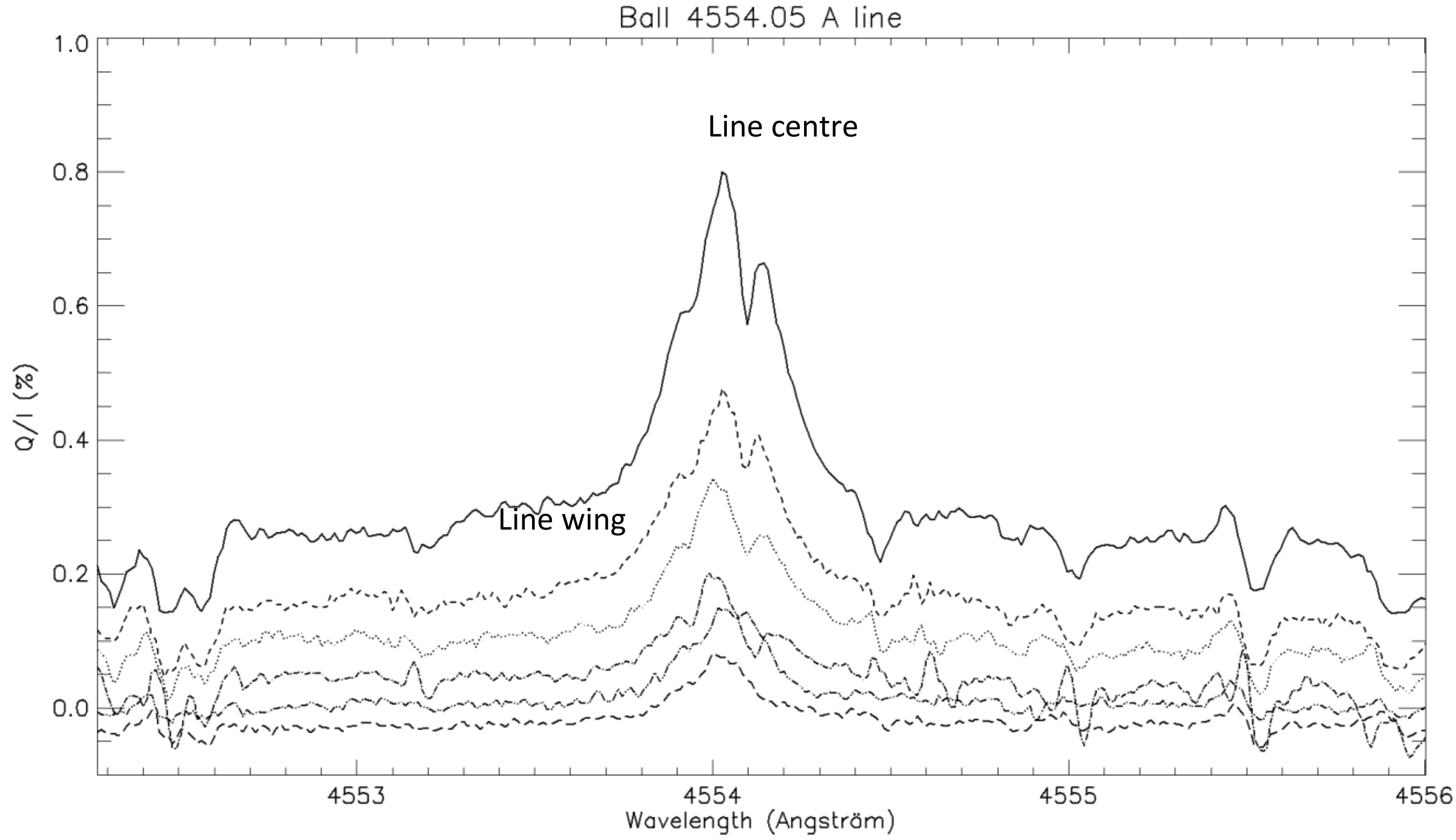

*Figure 7 : Q/I(λ) profiles (in %) at various limb distances of the limb (2", 5", 10", 20", 40", 80"). 15 May 2004, 10:20 – 11:20 UT.*

**III – 2 – SrI 4607 A (slit orthogonal to the limb near equator)**

Main results are shown in Figure 8 to 12; all scientific data are included in two datasets (two dates of observations with slit orthogonal to the limb). This line is a good candidate to probe weak magnetic fields with depth (Derouich *et al*, 2006).

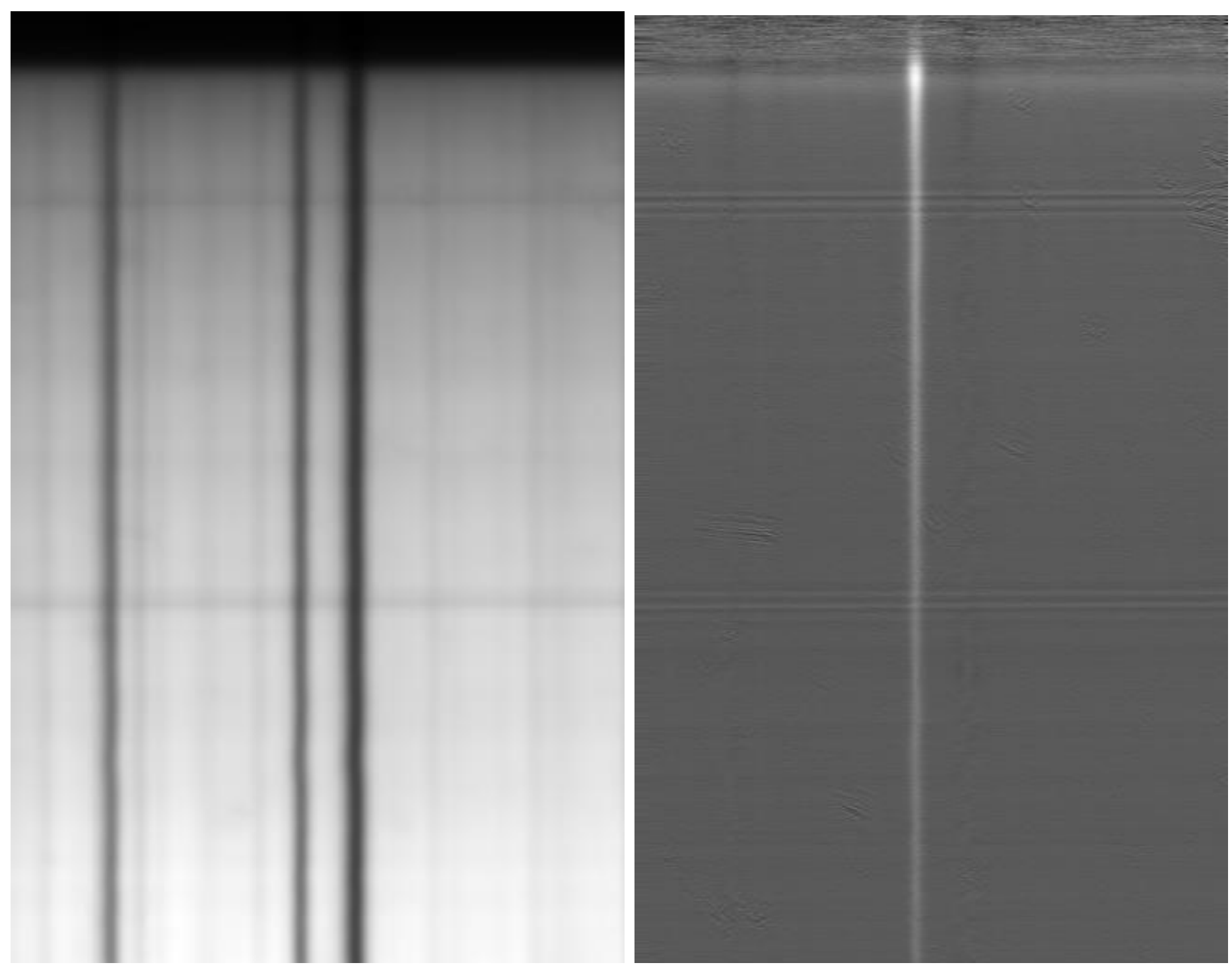

*Figure 8 : Intensity I(λ, x) and polarization rate Q/I(λ, x) where λ is the wavelength (in abscissa) and x (in ordinates) is the abscissa along the slit. 11 mA/pixel and 0.2"/pixel. 14 May 2004, 09:15 – 11:01 UT*

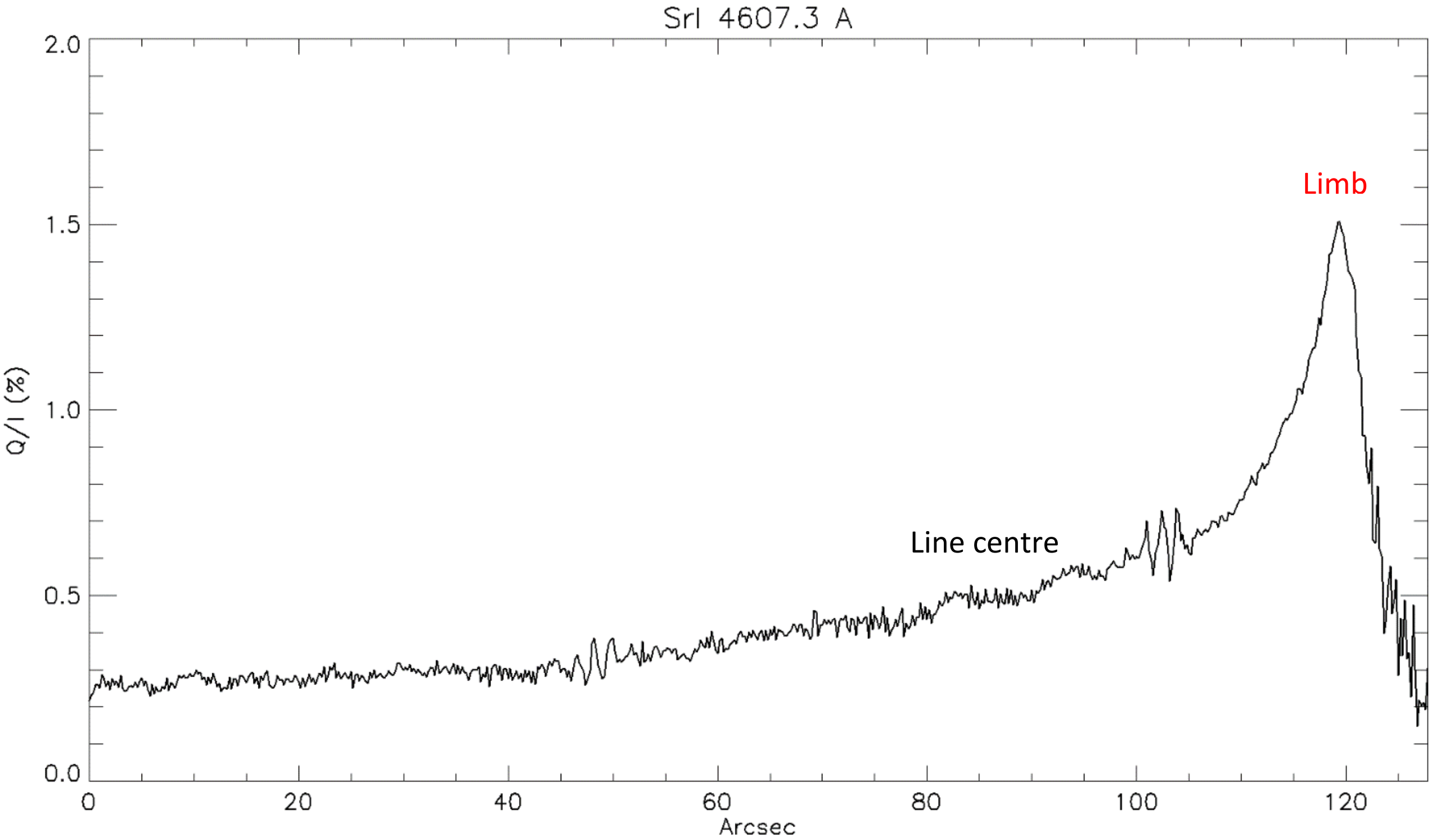

*Figure 9 : polarization rate Q/I along the slit (the limb is at the abscissa 120'') at line centre. 14 May 2004, 09:15 – 11:01 UT*

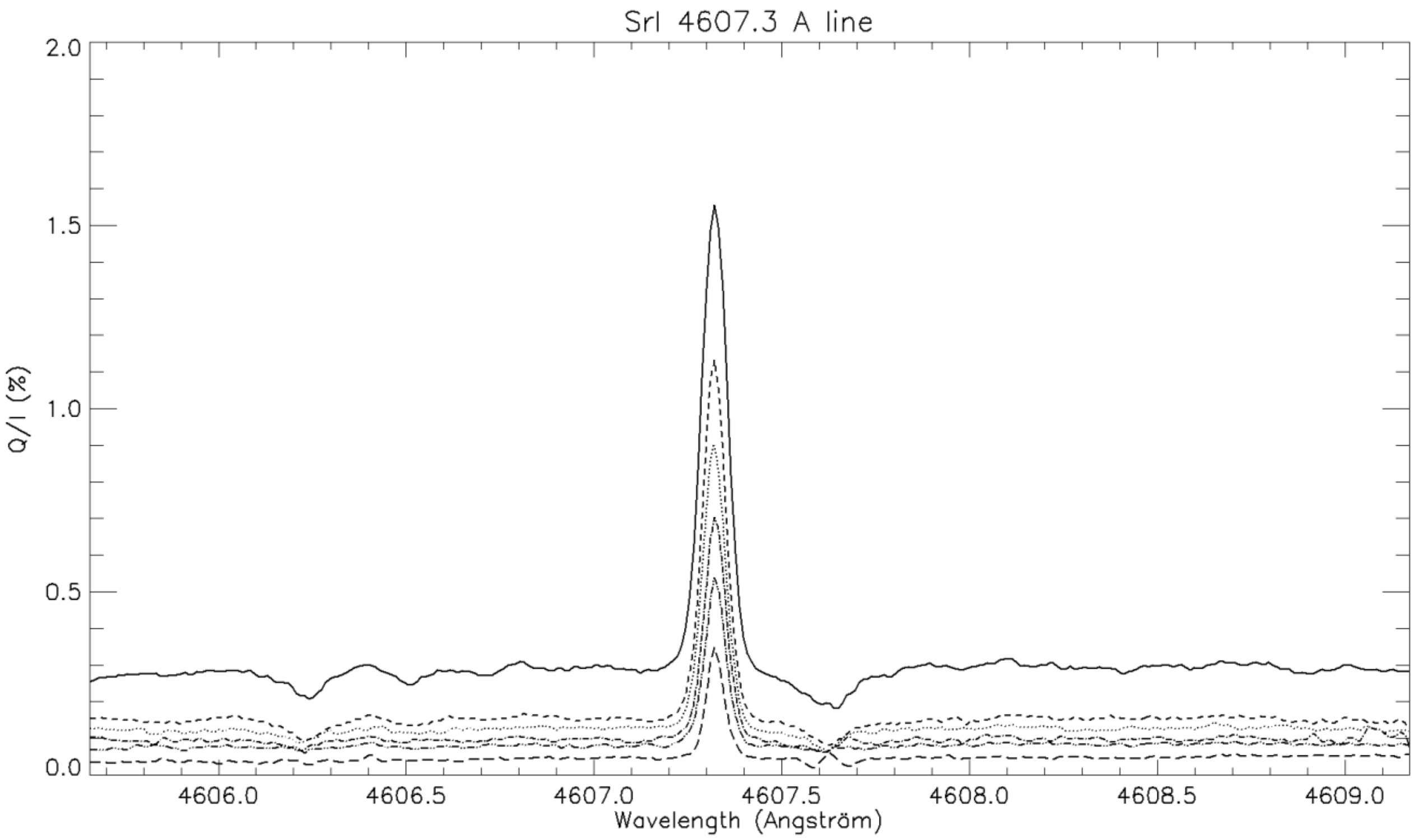

*Figure 10 : Q/I(λ) profiles (in %) at various limb distances (2'', 5'', 10'', 20'', 40'', 80'')*. 14 May 2004, 09:15 – 11:01 UT

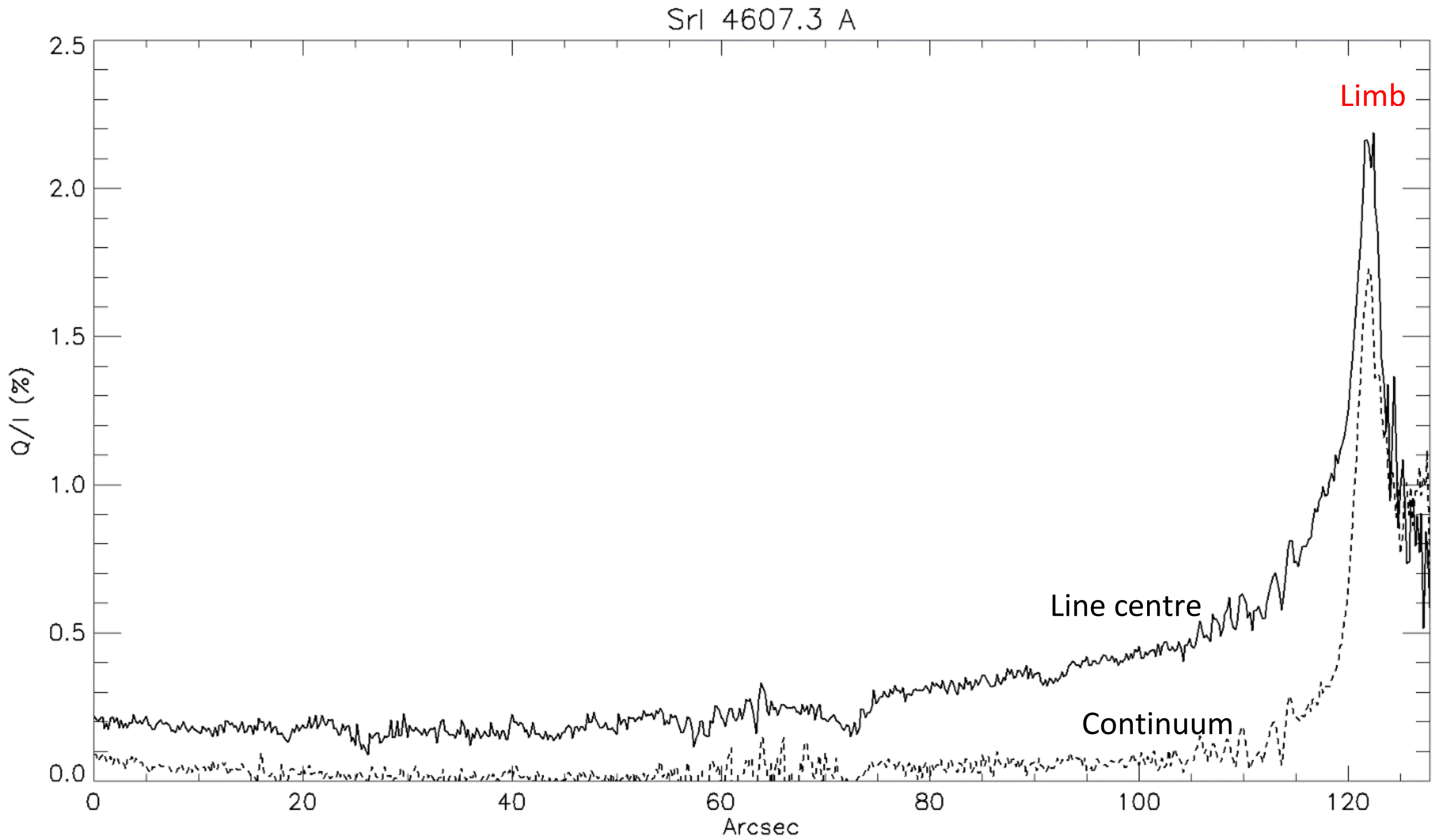

*Figure 11: polarization rate Q/I along the slit (the limb is at the abscissa 120”) at line centre and in the nearby continuum. 16 September 2004, 09:41 – 10:43 UT*

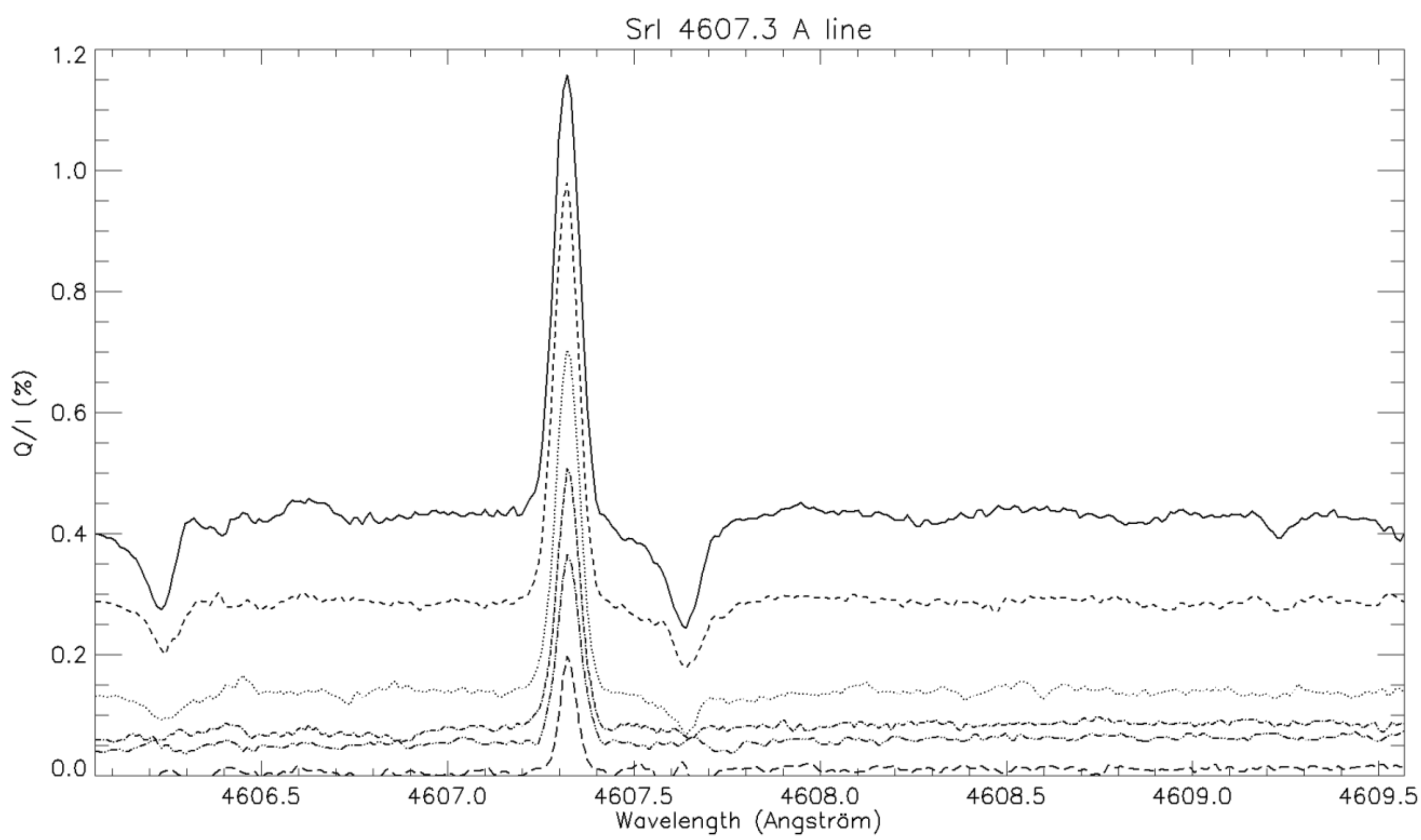

*Figure 12 : Q/I(λ) profiles (in %) at various limb distances (2”, 5”, 10”, 20”, 40”, 80”). 16 September 2004, 09:41 – 10:43 UT*

### III – 3 – SrI 4607 A (slit parallel to the limb near poles)

We performed also some observations of the polarization of SrI 4607 A (Figures 13 & 14) with the slit parallel to the limb, and at several discrete distances from the limb. In such a case, the integration time lasts 10 minutes instead of one hour or more, because it is possible to improve the SNR with the summation of all pixels along the slit (as in Figure 14).

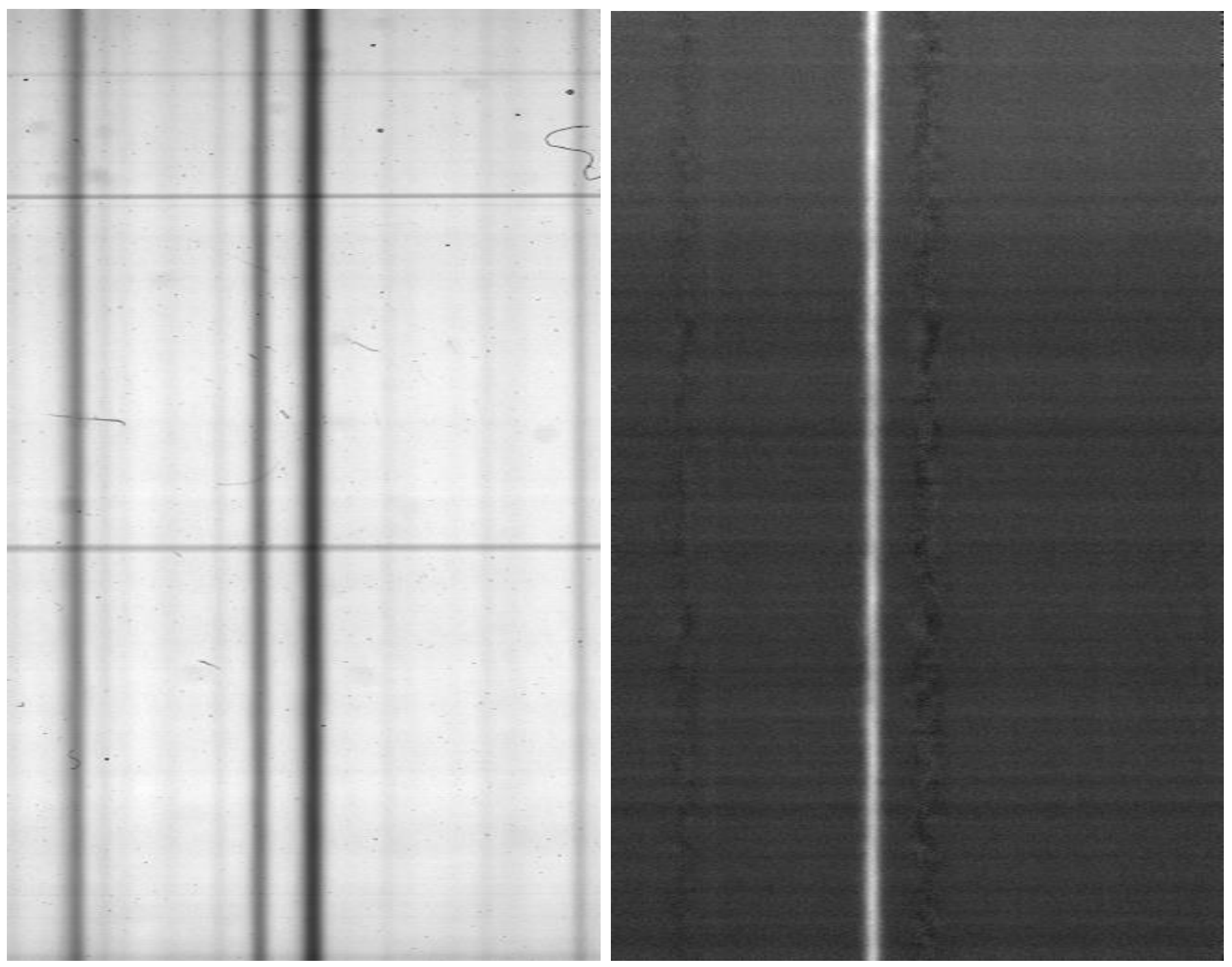

*Figure 13 : Intensity I(λ, x) and polarization rate Q/I(λ, x) where λ is the wavelength (in abscissa) and x (in ordinates) is the abscissa along the slit. 11 mA/pixel and 0.2"/pixel. Slit parallel to the limb at 5" distance. 15 May 2004, 06:22 – 06:32 UT*

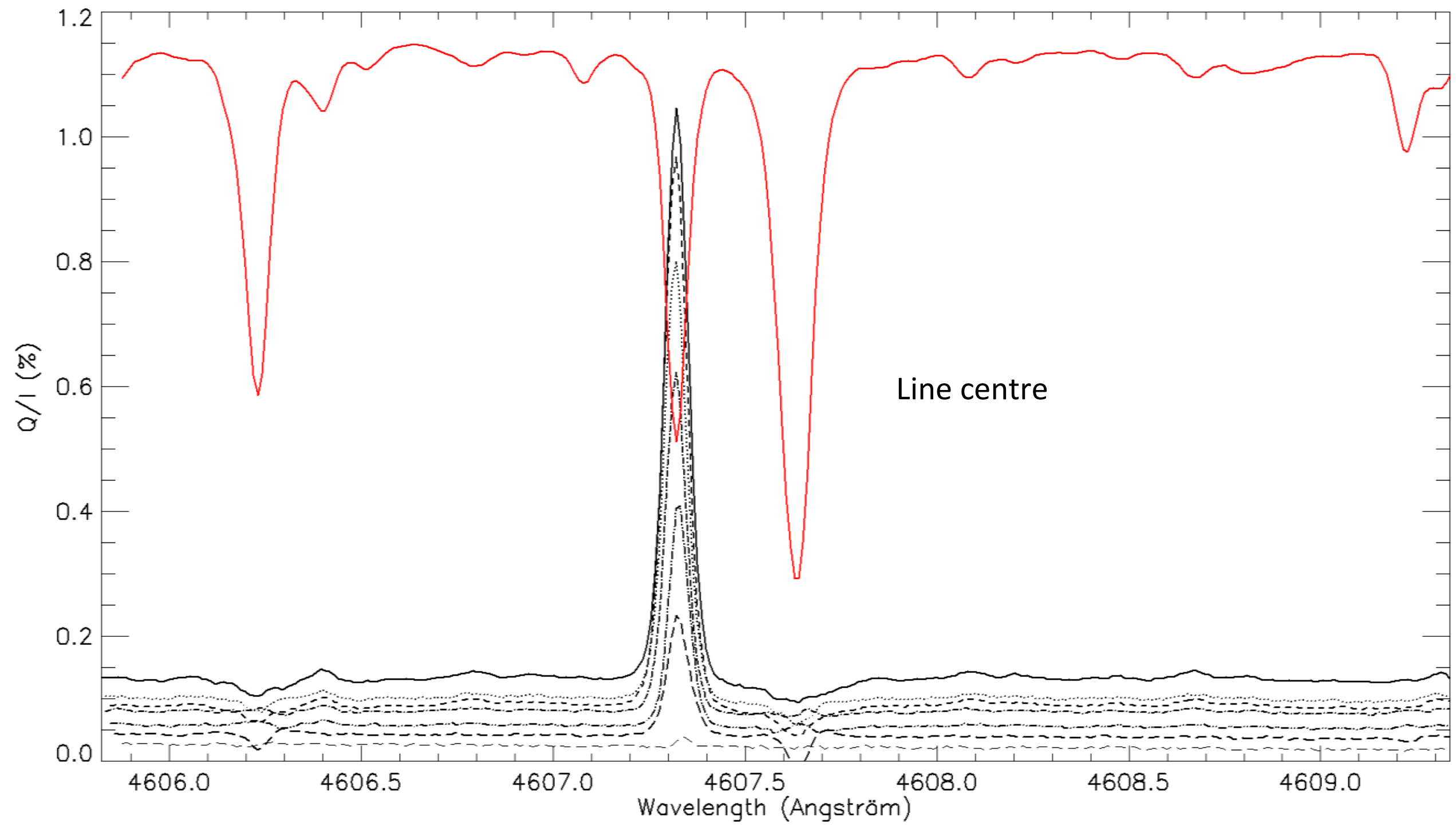

*Figure 14 : Q/I(λ) (in %) at various limb distances (2", 5", 10", 20", 40", 80", 160"). Intensity (in red) at 10" (the line depth will increase towards disk centre). 15 May 2004, 06:22 – 07:06 UT*

**III – 4 – SrII 4078 A (slit orthogonal to the limb near equator or poles)**

Main results are reported in Figures 15 to 20; all scientific FITS data, and more, are included in three datasets (two dates of observations with slit orthogonal to the limb near equator and one with slit parallel to the limb at 10" near poles). The core of this line seems very sensitive to Hanle depolarization by weak magnetic fields (Bianda *et al*, 1998), but this is not the case of wings which do not appear to vary so much, as shown by the figures.

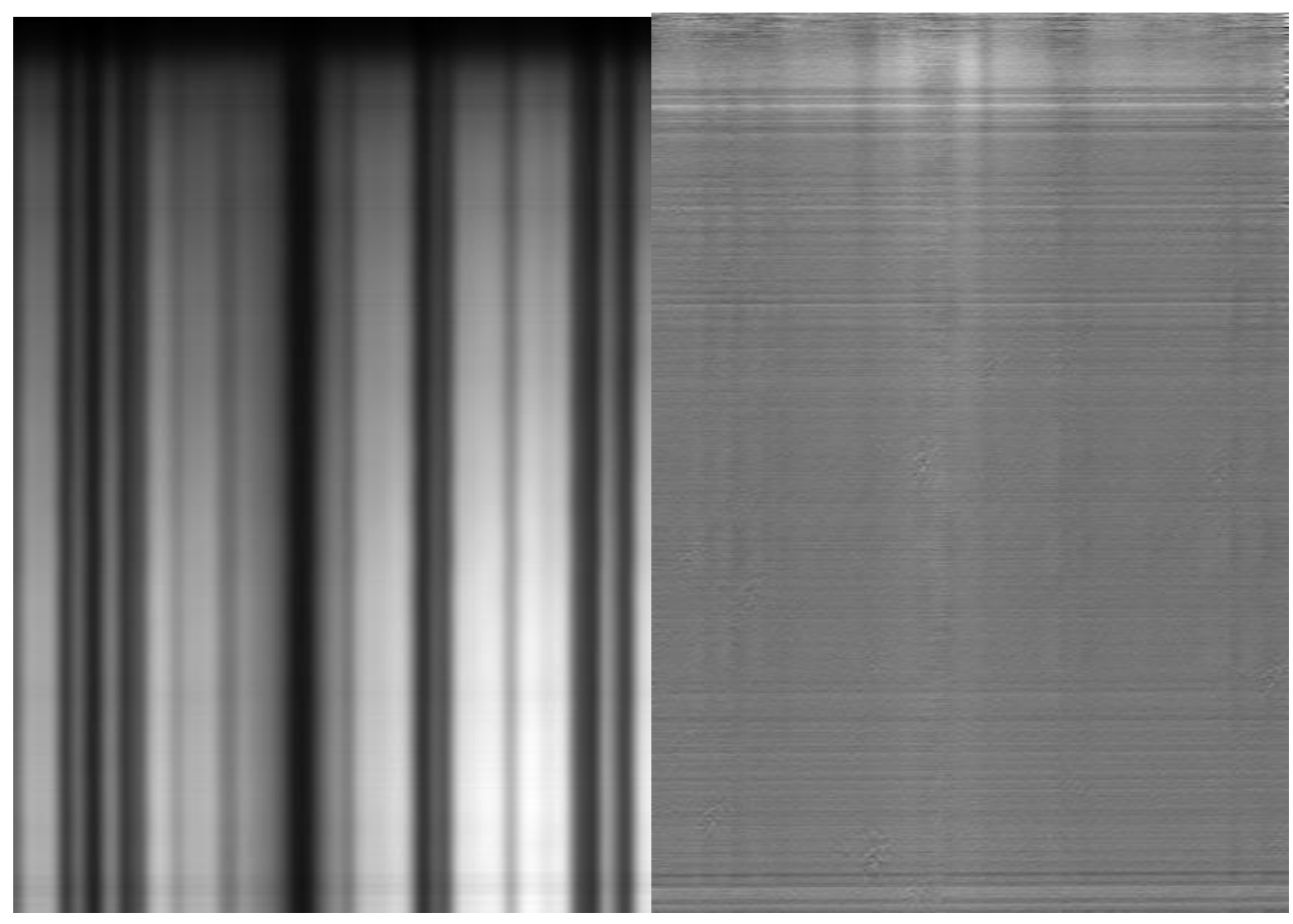

*Figure 15 : Intensity I(λ, x) and polarization rate Q/I(λ, x) where λ is the wavelength (in abscissa) and x (in ordinates) is the abscissa along the slit. 10 mA/pixel and 0.2"/pixel. 15 June 2006, 06:55 – 07:43 UT.*

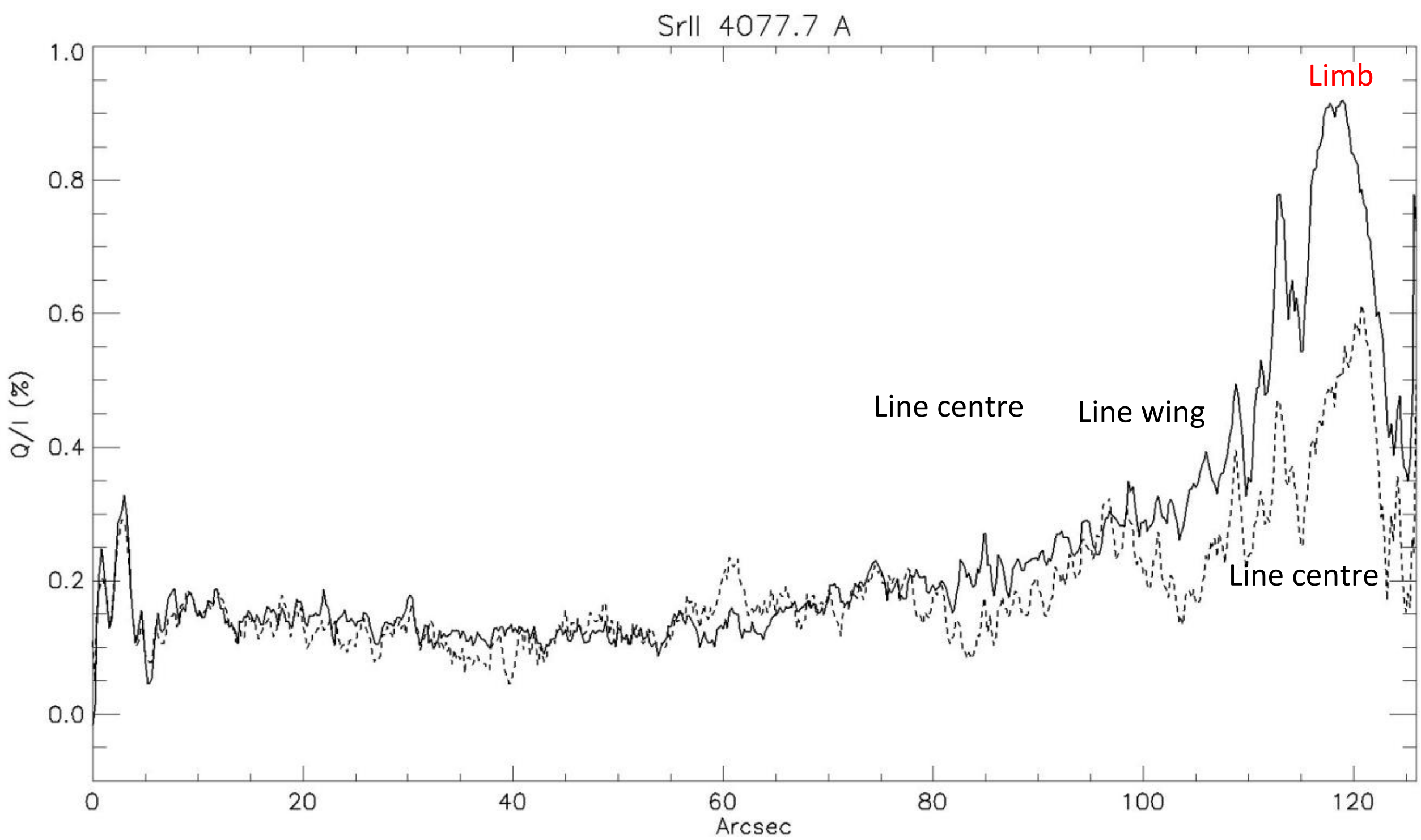

*Figure 16 : polarization rate Q/I along the slit (the limb is at the abscissa 120") at line centre and in the continuum. 15 June 2006, 06:55 – 07:43 UT.*

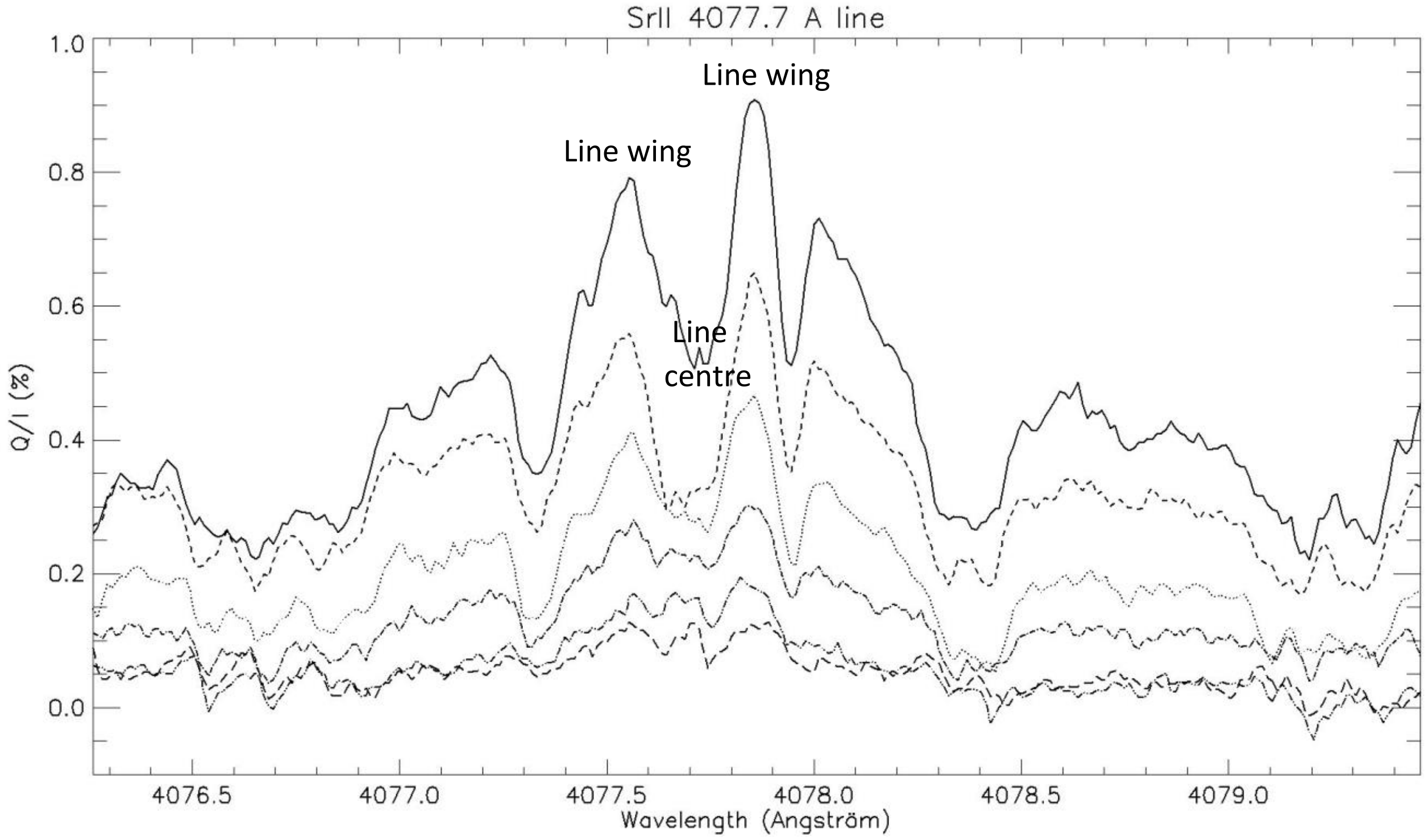

*Figure 17 : Q/I profiles (in %) as a function of wavelength at various limb distances (2", 5", 10", 20", 40", 80"). The polarization at line centre is reduced in comparison to both wings, it is not a systematic behaviour, as shown by Figure 20, this may be due to Hanle depolarization by turbulent magnetic fields, which are probable near equator. 15 June 2006, 06:55 – 07:43 UT.*

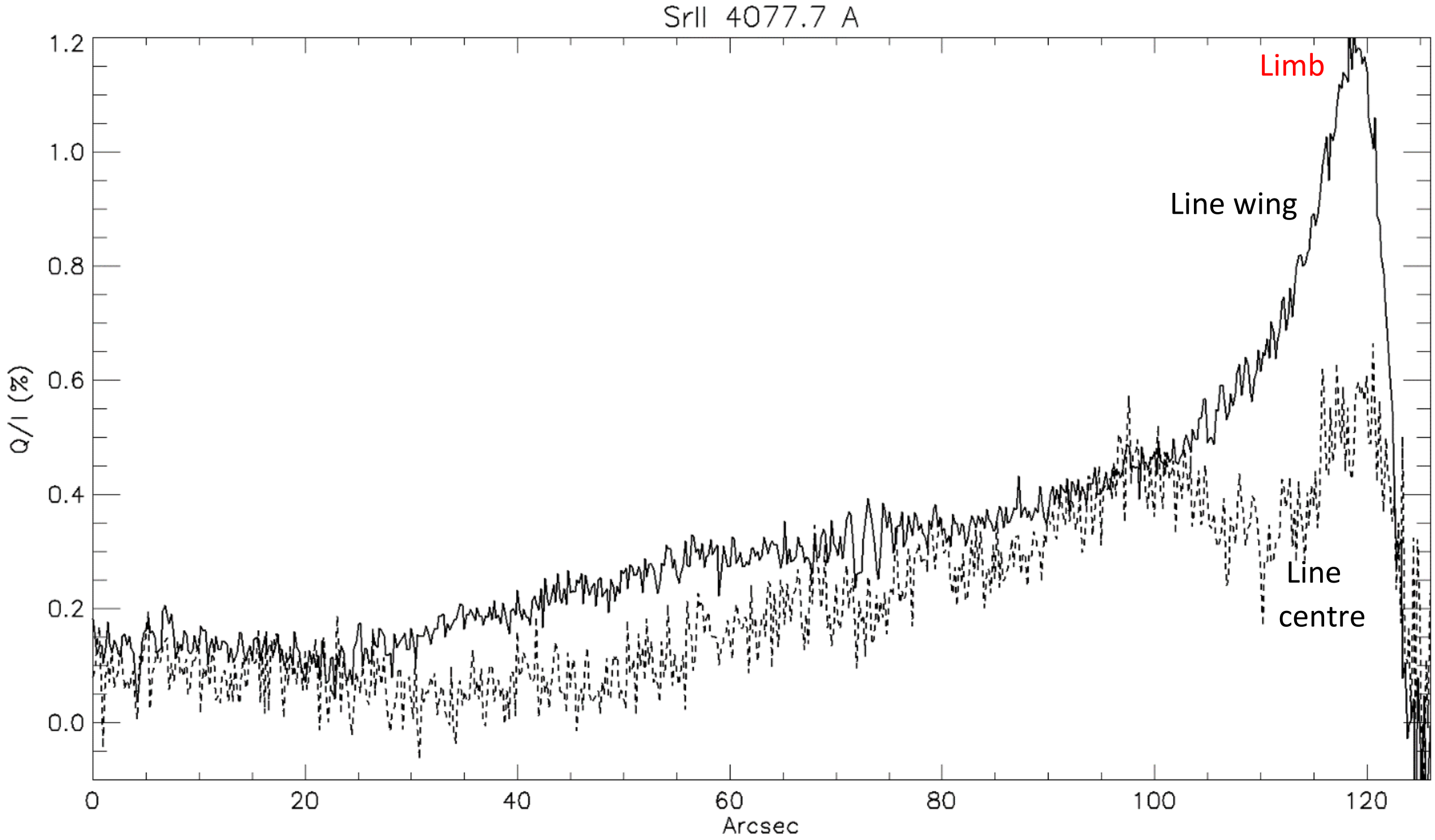

*Figure 18 : polarization rate Q/I along the slit (the limb is at the abscissa 120") at line centre and in the nearby continuum. 17 September 2004, 13:14 – 14:48 UT.*

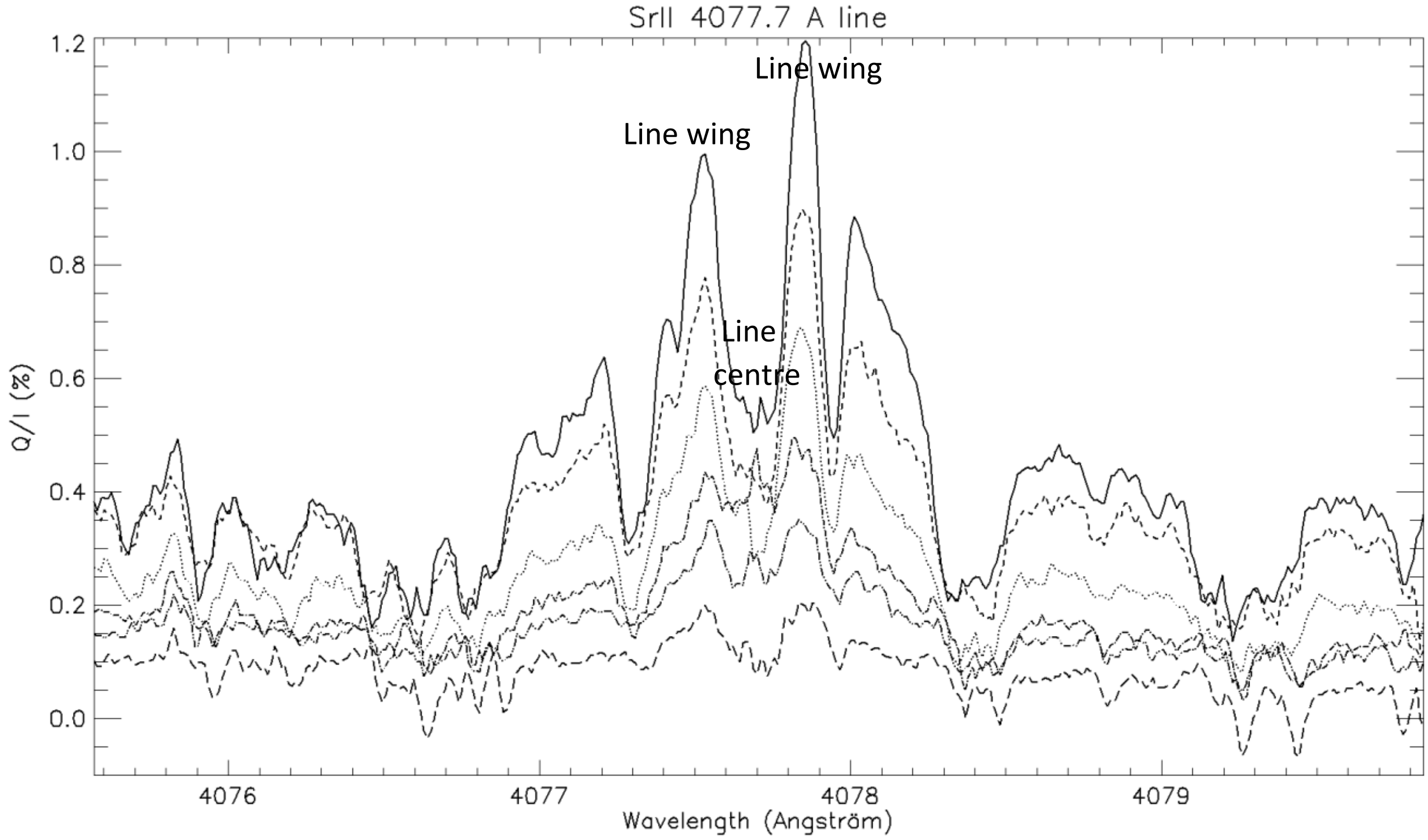

*Figure 19 : Q/I profiles (in %) as a function of wavelength at various limb distances (2", 5", 10", 20", 40", 80"). The polarization at line centre is reduced in comparison to both wings, it is not systematic, as shown by Figure 17, this may be due to Hanle depolarization by turbulent magnetic fields as observations occurred near equator. 17 September 2004, 13:14 – 14:48 UT.*

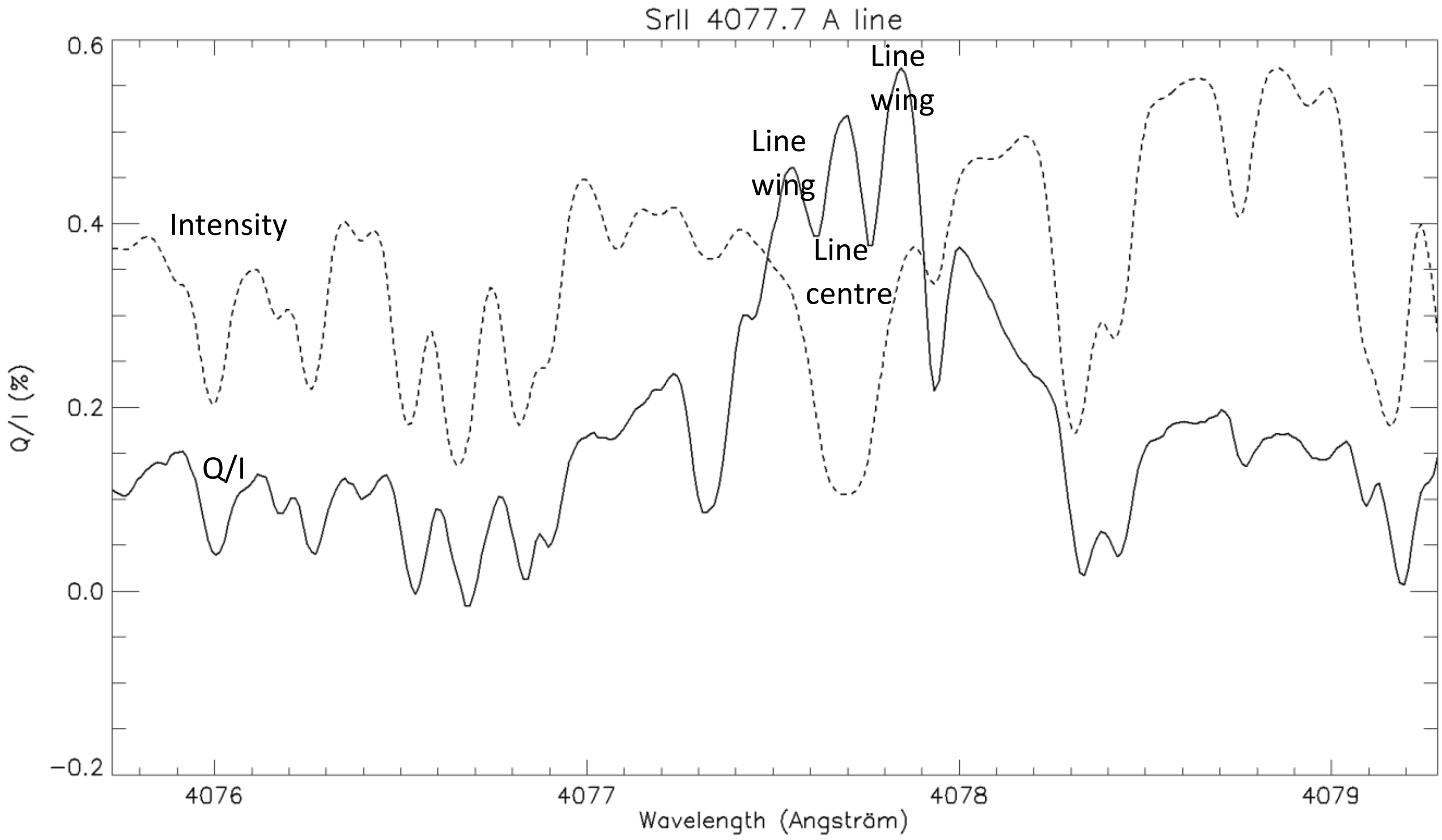

*Figure 20: Q/I profile (in %) as a function of wavelength at 10" limb distance. The polarization at line centre is well visible in this case, contrarily to Figures 17 & 19, probably because this observation occurred near poles. 15 May 2004, 12:43 – 12:52 UT*

**III – 5 – CaI 4227 A (slit orthogonal to the limb near equator)**

Main results are displayed in Figures 21 to 23; all scientific FITS data, and more, are included in the unique dataset. This line was intensively studied by Bianda *et al* (1999).

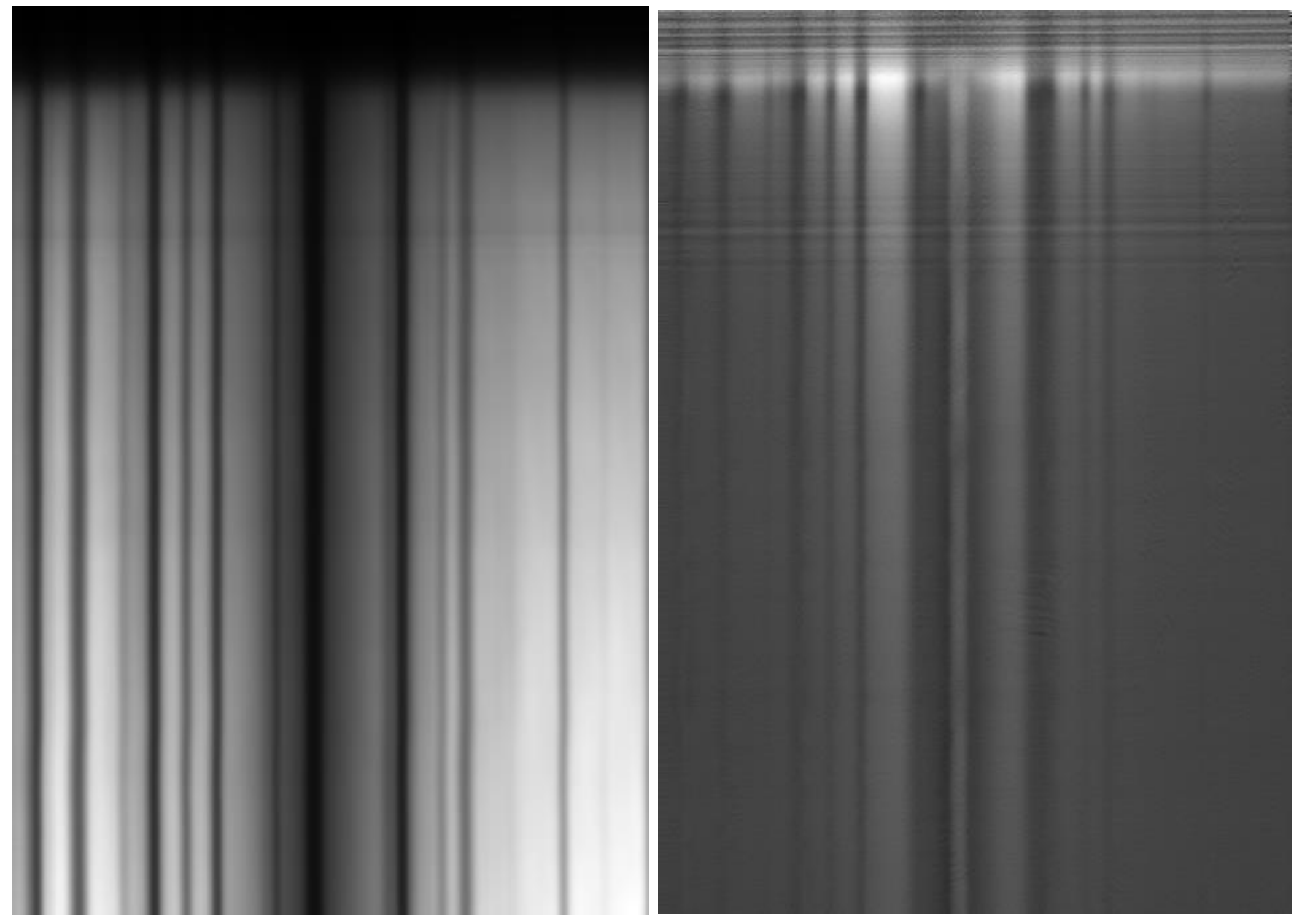

*Figure 21 : Intensity I(λ, x) and polarization rate Q/I(λ, x) where λ is the wavelength (in abscissa) and x (in ordinates) is the abscissa along the slit. 10 mA/pixel and 0.2''/pixel. 15 May 2004, 13:13 – 14:56 UT.*

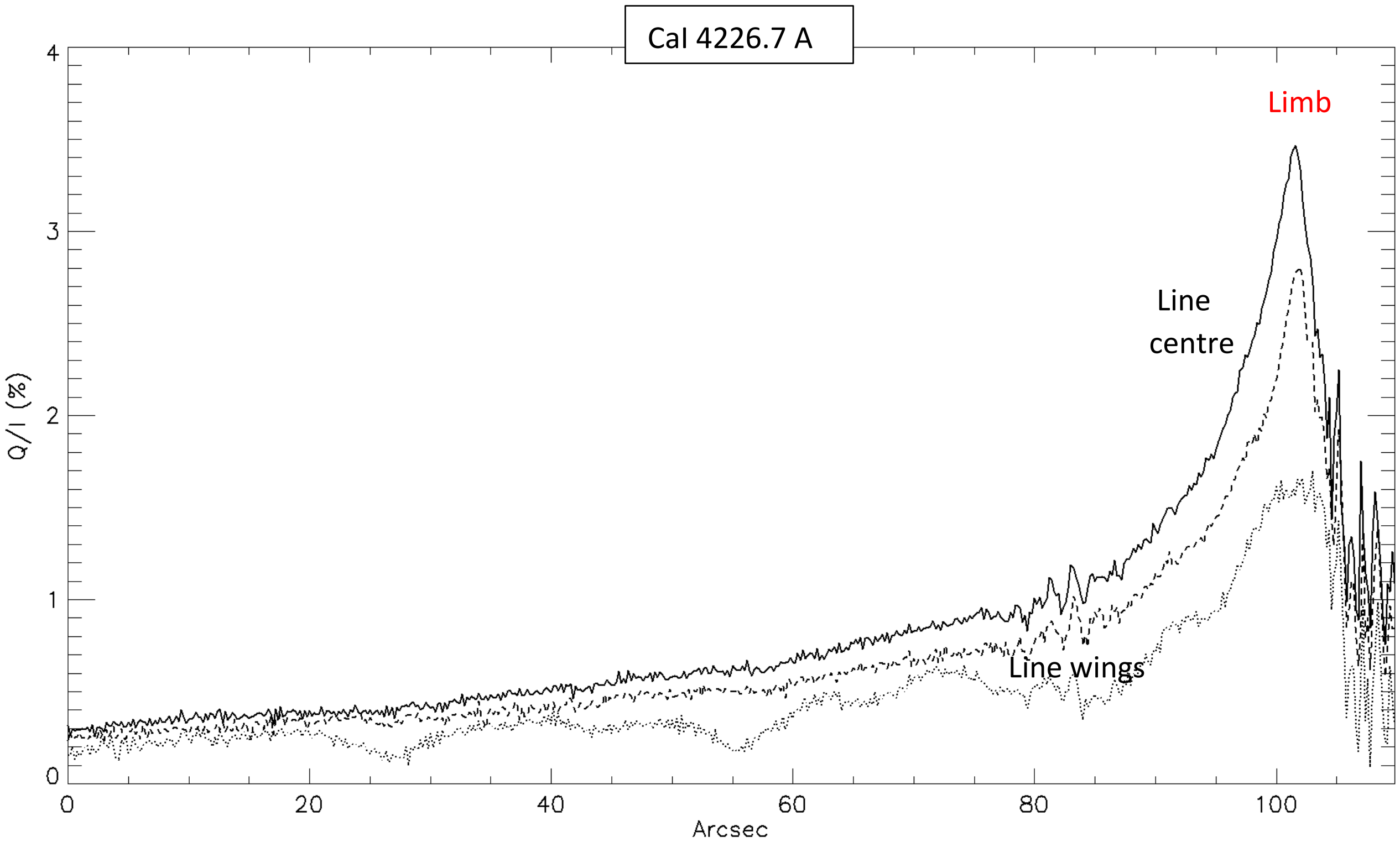

*Figure 22: polarization rate Q/I along the slit (the limb is at abscissa 100'') at line centre and both wings. 15 May 2004, 13:13 – 14:56 UT.*

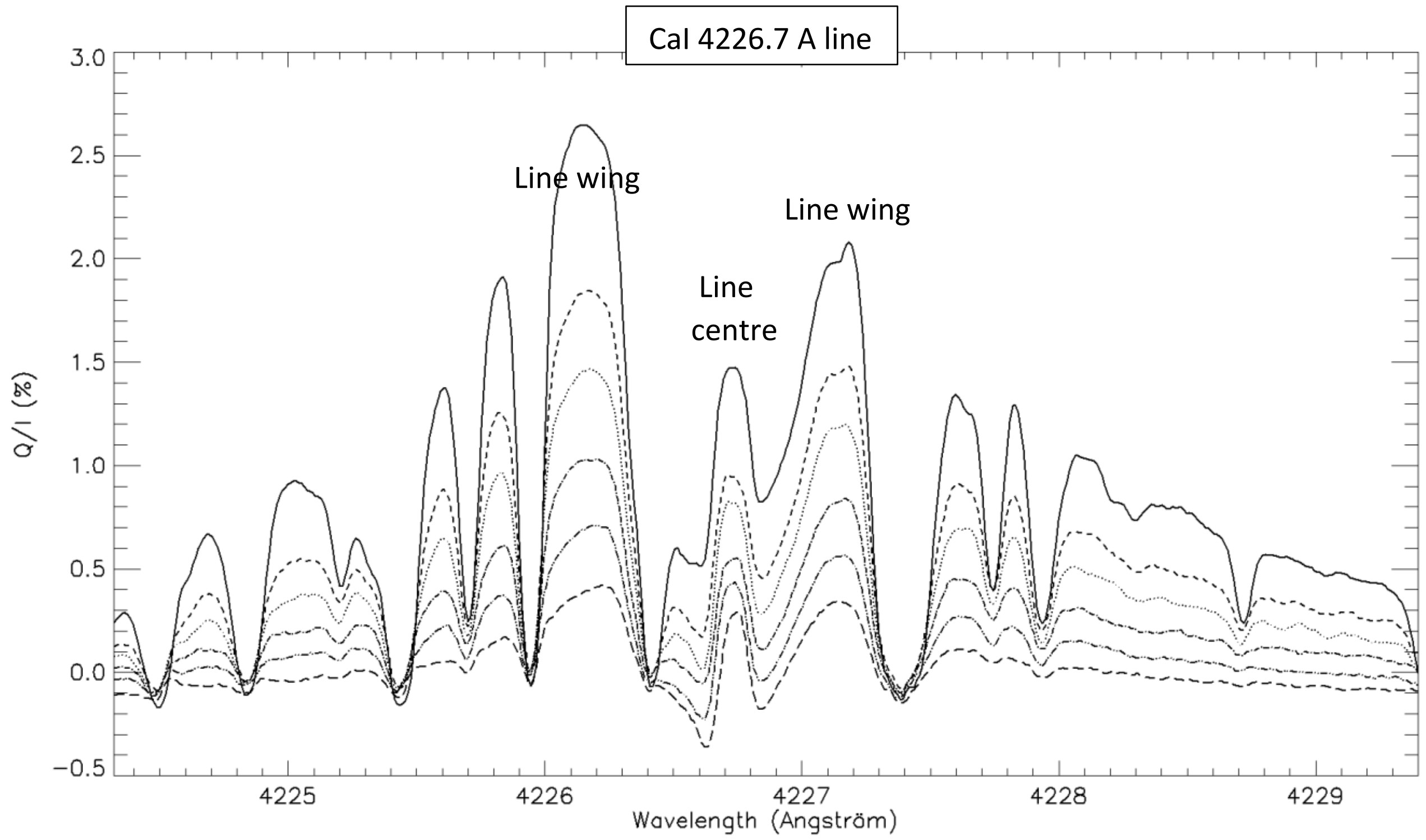

*Figure 23 : Q/I(λ) profiles (in %) at various limb distances (2", 5", 10", 20", 40", 80"). 15 May 2004, 13:13 – 14:56 UT.*

## IV – THE DATASET

The present data (and more) are available on line in scientific FITS format with a description at : https://entrepot.recherche.data.gouv.fr/dataset.xhtml?persistentId=doi:10.57745/DF54CS

## V - CONCLUSION

The present data were obtained with a liquid crystal polarimeter and a classical interline CCD camera running at 5-10 Hz cadence. Stokes combination I+Q and I-Q were obtained sequentially, 16 couples were integrated in real time within a few seconds to achieve the SNR of 500, in order to form two elementary 16 bit frames ; then tens or hundreds of such frames were got in sequence and added, in order to increase the SNR to 10000. Our data are not perfect and suffer, in some cases, from uncorrected flat fielding. However, observations with the slit orthogonal to the limb are not frequent and allow us to provide continuous curves of polarization from the limb to 80" distance below. As there was no field rotator, observations occurred near equator (except for a few cases with slit parallel to the limb near poles). Such data concerning CaI 4227 A, BaII 4554 A, SrI 4607 A and SrII 4078 A are of interest for measurements of weak unresolved magnetic fields and variations with depth, through the Hanle depolarization. It appeared clearly in the core of SrII 4078 between equator and poles.

## VI - REFERENCES